\documentclass[pra]{revtex4}
\newcommand{\be}{\begin{equation}}
\newcommand{\ee}{\end{equation}}
\newcommand{\bea}{\begin{eqnarray}}
\newcommand{\eea}{\end{eqnarray}}
\usepackage{epsfig}

\begin{document}

 \title{Vortex lattice of a Bose-Einstein Condensate in a rotating anisotropic trap}

\author{M.~\"O.~Oktel}
\affiliation{ Department of Physics, Bilkent University, 06800 Ankara, Turkey }%

\date{\today}

\begin{abstract}
We study the vortex lattices in a Bose-Einstein Condensate in a rotating anisotropic harmonic trap. We first investigate
the single particle wavefunctions obtained by the exact solution of the problem and give simple expressions
for these wavefunctions in the small anisotropy limit. Depending on the strength of the interactions, a few or
a large number of vortices can be formed. In the limit of many vortices, we calculate the density profile of the
cloud and show that the vortex lattice stays triangular. We also find that the vortex lattice planes align themselves with
the weak axis of the external potential. For a small number of vortices, we numerically solve the Gross-Pitaevskii equation
and find vortex configurations that are very different from the vortex configurations in an axisymmetric rotating trap.
\end{abstract}

\maketitle

\section{Introduction}
The study of rotating Bose-Einstein Condensates (BEC) have
progressed rapidly over the last three years. After the initial
demonstration of vortices at JILA and ENS \cite{mah99,mcw00}, regular lattices
containing hundreds of vortices have been formed \cite{arv01,hce01}. These vortices
were found to form remarkably regular triangular lattices. More
recently, the dynamical properties of BEC's containing vortex
lattices have been investigated \cite{ech02,ces03}.

The success of experimental groups in creating and probing vortex
structures have led to a flurry of theoretical activity. The
static and dynamical properties of a BEC with a vortex lattice \cite{ho01,mho03,cst03,bay03,bro99}, as
well as vortex lattices of spinor and multi component condensates \cite{mho02,kmm02,ktu02}
are active areas of theoretical research. Also, the presence of
multiply quantized vortices \cite{kba02,fet01} and melting of the vortex lattice \cite{cwg01,shm02} are
being discussed. The aim of theory is to clearly understand the
behavior of a BEC with a vortex lattice under different
conditions. The versatility of the dilute gas BEC experiments
enables theorists to consider different, experimentally realizable,
scenarios to discuss the response of the vortex lattice to various
external stimuli.

In this paper, we consider a BEC in a rotating anisotropic trap.
This system is different from most of the experimental realizations of
rotating BEC's to this date . In most  vortex lattice
experiments, and the theoretical works that try to explain them,
the rotating gas is first given angular momentum and then placed
in an axisymmetric trap, which conserves this angular momentum (for an exception see Ref.\cite{hhh02}). The
gas then rotates with an angular frequency that will minimize its
free energy. In this paper, however, we will assume that the BEC is
placed in an axially anisotropic trap that is rotating with a
fixed angular frequency $\Omega$, and is in its equilibrium state
in this trap. Theoretically a similar scenario was considered in Ref.\cite{lnf01,kav02}, however
both of these works considered vortex states with small number of vortices and did not present any calculations
pertaining to a large vortex lattice.

Recently such an anisotropic rotating potential was applied to a
BEC with a vortex lattice to induce collective oscillations \cite{ech02}. In
that experiment, the gas is rotating at a much higher angular
frequency than the applied anisotropic potential, thus the obtained results
 are not directly relevant to the theory
presented here. It is, however, important that a rotating
anisotropic trap was demonstrated experimentally and we believe
that the system we consider in this paper can be realized in a
similar experiment. An important result of that experiment was
that a vortex lattice that is not triangular was observed,
although it was not an equilibrium state. One is naturally led to
ask whether such a structural change in the lattice is possible as
an equilibrium state, when the anisotropic potential is static
in the rotating frame. The main result of this paper is that, when the cloud
contains a large number of vortices, the
deviation from the triangular lattice is small  and no structural phase transition should be expected. However, the configuration of
a small number of vortices can be markedly different from the
vortex configurations of an axisymmetric trap, in agreement with Ref.\cite{lnf01}. Also, an
anisotropic trap was used to study vortex generation \cite{hhh02},
and to investigate the escape of a BEC from the trap \cite{rps02}. The scenario considered in this paper  may be
experimentally realized in a setup  similar to one of these three experiments.

The paper is organized as follows. In section II, we find the
single particle wavefunctions for a rotating anisotropic trap by
exactly diagonalizing the Hamiltonian. We then show that the
wavefunctions in the lowest Landau level (LLL) can be written as simple
analytic functions in the small anisotropy limit. Sections III, IV
and V contain  discussions related to a condensate with a large
number of vortices. In section III, we give a variational
wavefunction for BEC with a vortex lattice in the lowest Landau
level of the anisotropic trap. We then use this wavefunction to calculate the density
profile of the cloud as a function of the rotation frequency and
the anisotropy of the trap. In section IV, we show that the vortex
lattice stays triangular and calculate the small distortion of the
lattice caused by the anisotropy of the density profile. In
section V, we show that the minimum energy configuration of the
lattice corresponds to lattice planes aligning with the weak axis
of the external potential. Section VI contains a discussion of the
vortex structures when only a small number of vortices are
present. We present the results of the numerical solution of the
Gross-Pitaevskii equation using the wavefunctions found in section
II. Finally, we give a summary of the results in section VII and
discuss their consequences for experiments.

\section{Single Particle Eigenstates}

To be able to discuss the properties of a BEC in a rotating anisotropic trap, we first consider a single
particle in such a trap. This problem is exactly solvable (see \cite{lnf01} and references therein) and
this solution guides us when we introduce a
variational wavefunction for the vortex lattice in the next section. Also, we use a truncated basis of
the exact single particle wavefunctions to study the BEC containing a small number of vortices in section VI.

We start by introducing the Hamiltonian for a two dimensional axisymmetric trap
\be
{\cal H}_0 = \frac{1}{2} m \omega_0^2 (x^2 + y^2) + \frac{1}{2 m}(p_x^2 + p_y^2).
\ee
An anisotropic trap will have an additional term
\be
{\cal H} = {\cal H}_0 + \frac{1}{2} m \Lambda^2 (x^2 - y^2).
\ee
In a reference frame rotating with an angular frequency $\Omega'$, the system is described by adding a term
$- \Omega' L$ to the Hamiltonian. Using lengths measured in units of oscillator length
\be
l = \sqrt{\frac{\hbar}{m \omega_0}},
\ee
and measuring energy in units of $\hbar \omega_0$, the Hamiltonian is
\be
\label{hamiltonian}
{\cal H} = \frac{1}{2} \left[ (1+\eta) x^2 + (1-\eta)y^2 +p_x^2 +p_y^2 \right] - \Omega ( x p_y - y p_x).
\ee
Here the dimensionless rotation frequency is
\be
\Omega = \frac{\Omega'}{\omega_0},
\ee
and the dimensionless anisotropy is
\be
\eta = \frac{\Lambda^2}{\omega_0^2}.
\ee
In the limit of $\eta \rightarrow 0$, we recover the Hamiltonian for a rotating axisymmetric trap.

This Hamiltonian can be diagonalized by the linear canonical transformation
\bea
u' &=& \gamma_x \left( \cos(\phi) x - \sin(\phi) p_y \right) \\ \nonumber
q'_u &=& \frac{1}{\gamma_x} \left( \sin(\phi) y + \cos(\phi) p_x \right) \\ \nonumber
v' &=& \gamma_y \left( \cos(\phi) y - \sin(\phi) p_x \right) \\ \nonumber
q'_v &=& \frac{1}{\gamma_y} \left( \sin(\phi) x + \cos(\phi) p_y \right),
\eea
where
\be
\tan(2 \phi) = \frac{2 \Omega}{\eta},
\ee
and
\bea
\label{gammas}
\gamma_x &=& \left[ \frac{ 1 + \frac{\eta}{2} + \sqrt{\Omega^2 + \left(\frac{\eta}{2}\right)^2 }}{ 1 - \frac{\eta}{2} + \sqrt{\Omega^2 + \left(\frac{\eta}{2}\right)^2}} \right]^{\frac{1}{4}}, \\ \nonumber
\gamma_y &=& \left[ \frac{ 1 - \frac{\eta}{2} - \sqrt{\Omega^2 + \left(\frac{\eta}{2}\right)^2 }}{ 1 + \frac{\eta}{2} - \sqrt{\Omega^2 + \left(\frac{\eta}{2}\right)^2}} \right] ^{\frac{1}{4}}.
\eea
The diagonal Hamiltonian is then
\bea
{\cal H} &=& \frac{1}{2} \sqrt{ 1 + 2 \sqrt{\Omega^2 + \left( \frac{\eta}{2} \right)^2} + \Omega^2 } \left( u'^2 +q'^2_u \right) \\ \nonumber
&+& \frac{1}{2} \sqrt{ 1 - 2 \sqrt{\Omega^2 + \left( \frac{\eta}{2} \right)^2} + \Omega^2 } \left( v'^2 +q'^2_v \right).
\eea

As $u',q'_u$ and $v',q'_v$ are canonically conjugate pairs, the energy spectrum in units of $\hbar \omega_0$ is simply
\be
\epsilon = n \epsilon_u + m \epsilon_v + \frac{1}{2} (\epsilon_u+\epsilon_v),
\ee
where
\bea
\epsilon_u = \sqrt{ 1 + 2 \sqrt{\Omega^2 + \left( \frac{\eta}{2} \right)^2} + \Omega^2 } \\
\epsilon_v = \sqrt{ 1 - 2 \sqrt{\Omega^2 + \left( \frac{\eta}{2} \right)^2} + \Omega^2 },
\eea
and $n$, $m$ are non-negative integers. To get a spectrum that is bounded from below, we must have
\be
\label{criticalcondition}
\eta \le 1 - \Omega^2.
\ee
If this condition is not satisfied, the centrifugal force overcomes the trapping potential in the weak
direction and the particle will no longer be trapped. However, when this condition is satisfied as an
equality, $\epsilon_v$ is zero and there is an infinite set of degenerate eigenstates forming the analogue
of the lowest Landau level (LLL) \cite{kmp00} for an anisotropic trap. The formation of such a Landau level will
simplify the discussion of the vortex lattice considerably.

To obtain a useful form for the eigenstates of the anisotropic trap in real space, we express the
creation-annihilation operators of the anistropic rotating trap in terms of the creation-annihilation
operators of the the axisymmetric rotating trap. For the axisymmetric trap we define
\be
a_{0u}=\frac{1}{\sqrt{2}}(a_x + i a_y) \hspace{1cm} a_{0v}= \frac{1}{\sqrt{2}}(a_y + i a_x).
\ee
Similarly, for the anisotropic trap, we have
\be
a_u = \frac{1}{\sqrt{2}}(u' + i q'_u) \hspace{1cm} a_v = \frac{1}{\sqrt{2}}(v' + i q'_v).
\ee
We can express the latter in terms of the former as
\bea
\label{operatortransformation}
a_u &=& \frac{1}{2} \left[ (\gamma_x + \frac{1}{\gamma_x}) a_1 + (\gamma_x - \frac{1}{\gamma_x}) a_1^\dagger \right] \\ \nonumber
a_v &=& \frac{1}{2} \left[ (\gamma_y + \frac{1}{\gamma_y}) a_2 + (\gamma_y - \frac{1}{\gamma_y}) a_2^\dagger \right],
\eea
with
\bea
a_1 &=& \frac{1}{\sqrt{2}} \left[ (c+s) a_{0u} + i (s-c) a_{0v} \right],  \\ \nonumber
a_2 &=& \frac{1}{\sqrt{2}} \left[ (c+s) a_{0v} + i (s-c) a_{0u} \right].
\eea
The ground state of the axisymmetric Hamiltonian, $|00\rangle_0$, satisfies
\be
a_{0u} |00\rangle_0 = 0 \hspace{1cm} a_{0v} |00\rangle_0 = 0.
\ee
This state is also annihilated by $a_1$,$a_2$
\be
a_1 |00\rangle_0 = 0 \hspace{1cm} a_2 |00\rangle_0 = 0.
\ee
Similarly, the ground state of the anisotropic rotating trap, $|00\rangle$ is defined by
\be
a_u |00\rangle = 0 \hspace{1cm} a_v |00 \rangle =0.
\ee

Rewriting the definition of $a_u$ (\ref{operatortransformation}), we have
\be
a_u =  \frac{1}{2} \left[ (\gamma_x + \frac{1}{\gamma_x}) a_1 + (\gamma_x - \frac{1}{\gamma_x}) a_1^\dagger \right] = \frac{(\gamma_x - \frac{1}{\gamma_x})}{2} e^{-\frac{1}{2} \Theta_x (a_1^\dagger)^2} a_1  e^{\frac{1}{2} \Theta_x (a_1^\dagger)^2},
\ee
with
\bea
\Theta_x &=& \frac{\displaystyle{(\gamma_x - \frac{1}{\gamma_x})}}{\displaystyle{(\gamma_x + \frac{1}{\gamma_x})}} \\ \nonumber
\Theta_y &=& \frac{\displaystyle{(\gamma_y - \frac{1}{\gamma_y})}}{\displaystyle{(\gamma_y + \frac{1}{\gamma_y})}}.
\eea
Which gives us an easy way to identify the ground state of the anisotropic potential
\be
0 = a_u |0 \rangle = {2} e^{-\frac{1}{2} \Theta_x (a_1^\dagger)^2} a_1 \underbrace{ e^{\frac{1}{2} \Theta_x (a_1^\dagger)^2} |0 \rangle}_{ \displaystyle{ |0\rangle_0}}.
\ee
By reexpressing both $a_v$ and $a_u$ as above we obtain the correctly normalized ground state as
\be
\label{groundstate}
|00\rangle = \frac{2}{\sqrt{\displaystyle{(\gamma_x + \frac{1}{\gamma_x})(\gamma_y + \frac{1}{\gamma_y})}}} \: e^{\displaystyle{ -\frac{1}{2} \Theta_x (a_1^\dagger)^2}} \: e^{\displaystyle{ -\frac{1}{2} \Theta_y (a_2^\dagger)^2}} |00\rangle_0.
\ee
Higher states in the anisotropic rotating trap are
\be
|n m \rangle = \frac{(a_u^\dagger)^n}{\sqrt{n!}} \frac{(a_u^\dagger)^n}{\sqrt{n!}} |0 0 \rangle,
\ee
and their overlap with the ground state of the axisymmetric trap can be calculated as
\be
{}_0\langle 0 0 | n m \rangle = \frac{1}{(\frac{n}{2})! (\frac{m}{2})!} \sqrt{\frac{n! m!}{\displaystyle{2^{n-1} 2^{m-1}}}}  \sqrt{ \frac{\Theta_x^n \: \Theta_y^m}{\displaystyle{(\gamma_x + \frac{1}{\gamma_x})(\gamma_y + \frac{1}{\gamma_y})}}}.
\ee
These overlaps will help us represent the wavefunctions of the anisotropic potential in a simple form.

For the axisymmetric rotating trap, wavefunctions in the LLL are analytic functions of $z=x+iy$. When
the rotation frequency is close to the trapping frequency, these wavefunctions become the only states that
are populated due to the high energy cost of the higher Landau levels. These wavefunctions
have successfully been used to describe the properties of the vortex lattice, even if the experimental situation
does not correspond to the fast rotation limit \cite{ho01,mho02,mho03}. The analytical properties of a many-body wavefunction formed in the
LLL describes vortex formation very well, thus such a wavefunction can always be  used as a variational
wavefunction at any rotation frequency. Also, very recently, a BEC that is in the LLL regime was experimentally realized
\cite{sce03}.

For an anisotropic rotating trap, if the rotation frequency is close to the critical frequency defined by
Eq.(\ref{criticalcondition}), Landau levels are again formed and we can consider  only the wavefunctions in the
LLL to describe vortices. However, there is no simple expression for the wavefunctions in the LLL of the anisotropic
trap except in the limit,
\bea
\delta &=& 1- \Omega  \ll 1 \\ \nonumber
\eta &\ll& 1 \\ \nonumber
\frac{\eta}{2 \delta} &=& \lambda \simeq 1.
\eea
This limit corresponds to small anisotropy of the trap and fast rotation, and is as much experimentally accessible as
the fast rotation limit of the axisymmetric trap \cite{ho01,sce03}. It is also important to realize that, even if the anisotropy of the
trap is small, as the rotation frequency approaches the critical frequency, the cloud of atoms will be stretched
along the weak axis. This enables us to use the wavefunctions found in the small anisotropy limit to describe a BEC
with any aspect ratio in the plane of rotation. For a trap with large anisotropy such wavefunctions should only be
regarded as variational wavefunctions, while they should describe the system very well in the small anisotropy limit.

Thus, we concentrate on the small anisotropy limit and find an expression for the real space representation of the
wavefunctions in the LLL. In this limit various parameters in the exact solution reduce to
\bea
\cos(\phi) &=& \sin(\phi) = \frac{1}{\sqrt{2}} \\ \nonumber
\gamma_x &=& 1 \rightarrow \theta_x = 0 \\ \nonumber
\gamma_y &=& \left[ \frac{1-\frac{\lambda}{2}}{1+\frac{\lambda}{2}} \right]^{\frac{1}{4}} \\ \nonumber
\theta_y &=& \frac{ \left(1-\frac{\lambda}{2} \right)^{\frac{1}{2}} - \left(1+\frac{\lambda}{2} \right)^{\frac{1}{2}}}
{ \left(1-\frac{\lambda}{2} \right)^{\frac{1}{2}} + \left(1+\frac{\lambda}{2} \right)^{\frac{1}{2} }}.
\eea

Using these parameters, we calculate the spatial wavefunctions. Recalling that the ground state of the axisymmetric
trap is
\be
\label{isotropicgroundstate}
\langle xy | 00\rangle_0 = \frac{1}{\sqrt{\pi}} e^{-z \bar{z}},
\ee
with
\be
z=(x+i y)/\sqrt{2}.
\ee
The operators take the form
\bea
a_{0u}&=&(z+\partial_{\bar{z}})/\sqrt{2} \\ \nonumber
a_{0v}&=&i (\bar{z}+ \partial_{z})/\sqrt{2}.
\eea
By using (\ref{isotropicgroundstate}) we have
\be
\langle xy | 00 \rangle = \sqrt{\frac{2}{\gamma_y+\gamma_y^{-1}}} \frac{1}{\sqrt{\pi}} e^{\theta_y z^2- z \bar{z}}.
\ee
We can also find the form of the excited states as in this limit
\be
a_{v}^\dagger = \frac{\gamma_y+\gamma_y^{-1}}{2} ( a_{0v}^\dagger + \theta_y a_{0v}).
\ee
We have then
\be
\langle xy|0n\rangle = \frac{1}{\sqrt{\pi n!}} \sqrt{\frac{2}{\gamma_y+\gamma_y^{-1}}}
\left[\frac{-i (\gamma_y+\gamma_y^{-1})}{2 \sqrt{2}} \right]^n
\left[ z - \theta_y \bar{z} - (\partial_{\bar{z}}+\theta_y \partial_z)\right]^n
e^{\theta_y z^2 - z \bar{z}}.
\ee
We define
\be
I_n = \left[ z - \theta_y \bar{z} - (\partial_{\bar{z}}+\theta_y \partial_z)\right]^n
e^{\theta_y z^2 - z \bar{z}},
\ee
and use the transformation
\be
w = z - \theta_y \bar{z},
\ee
to calculate $I_n$ as
\be
I_n = (1-\theta_y^2)^{\frac{n}{2}} \theta_y^{\frac{n}{2}} H_n ( \sqrt{\theta_y^{-1} - \theta_y} z)
e^{\theta_y z^2 - z \bar{z}}.
\ee
Where $H_n$ are the Hermite polynomials.

Thus the wavefunctions in the lowest Landau level are given by
\be
\label{anisotropicwavefunctions}
\Psi_n = \frac{1}{\sqrt{ \pi n!}} \frac{(-i)^n}{2^{\frac{n-1}{2}}} (\gamma_y+\gamma_y^{-1})^{\frac{1}{2}}
\theta_y^{\frac{n}{2}} H_n\left( \frac{2 z}{\sqrt{\gamma_y^2 - \gamma_y^{-2}}} \right) e^{\theta_y z^2 - z \bar{z}}.
\ee
We check the orthogonality of these wavefunctions using a method which demonstrates how matrix elements of operators
can be calculated in this basis.  The overlap of two wavefunctions $|\Psi_n\rangle$ and $|\Psi_m \rangle$ is proportional
to
\be
\langle \Psi_n | \Psi_m \rangle \propto \int dx dy e^{-2 z \bar{z} + \theta_y ( z^2+\bar{z}^2) } H_n(\alpha \bar{z})
H_m(\alpha z),
\ee
with
\be
\alpha = \sqrt{\theta_y^{-1} - \theta_y}.
\ee
We can use the generating functions for Hermite polynomials
\be
H_n(\alpha z) = \left[\frac{d}{dt} e^{-t^2+2 t \alpha z} \right]_{t=0},
\ee
and then evaluate the gaussian integral. Which gives us
\be
\langle \Psi_n | \Psi_m \rangle \propto \left[ \frac{d^n}{dt^n} \frac{d^m}{ds^m} e^{\frac{2}{\theta_y} s t} \right]_{t=s=0}.
\ee
This expression is equal to zero unless $n=m$. With the correctly normalized wavefunctions, we obtain the orthonormality
relation
\be
\langle \Psi_n | \Psi_m \rangle = \delta_{n,m}.
\ee

In the small anisotropy limit presented here, we see that the anisotropic single particle wavefunctions are still analytic
functions of $z$, multiplied by a Gaussian. Thus, they only have overlaps with the LLL of the axisymmetric rotating trap.
The introduction of a small anisotropy reorganizes the LLL within itself, {\it i.e.} no wavefunctions from higher Landau
levels of the axisymmetric trap are mixed.

The n$^{th}$ single particle wavefunction of the axisymmetric trap corresponds to a state that has a single multiply quantized
vortex $z^n$ at the origin. For the anisotropic trap, however, the n$^{th}$ single particle wavefunction will have $n$ singly
quantized vortices that are distributed on the weak axis of the potential. Positions of these vortices are given by the
zeros of the Hermite polynomial $H_n$ (See Fig[1]).

\begin{figure}
\label{5statefig}
\centerline{\psfig{file=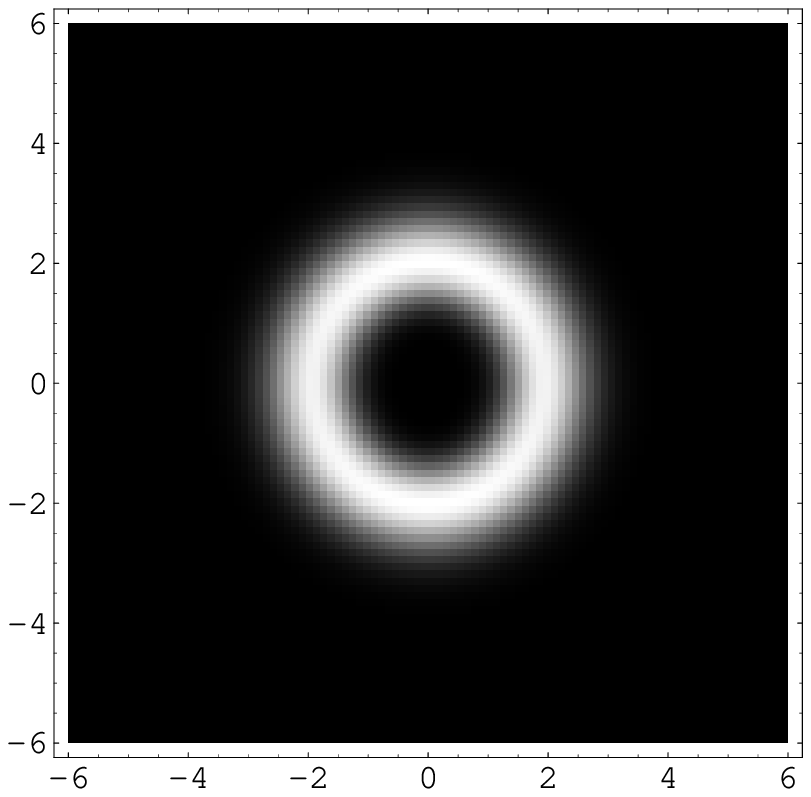,width=2in,height=2in}
\hspace{.25cm}
\psfig{file=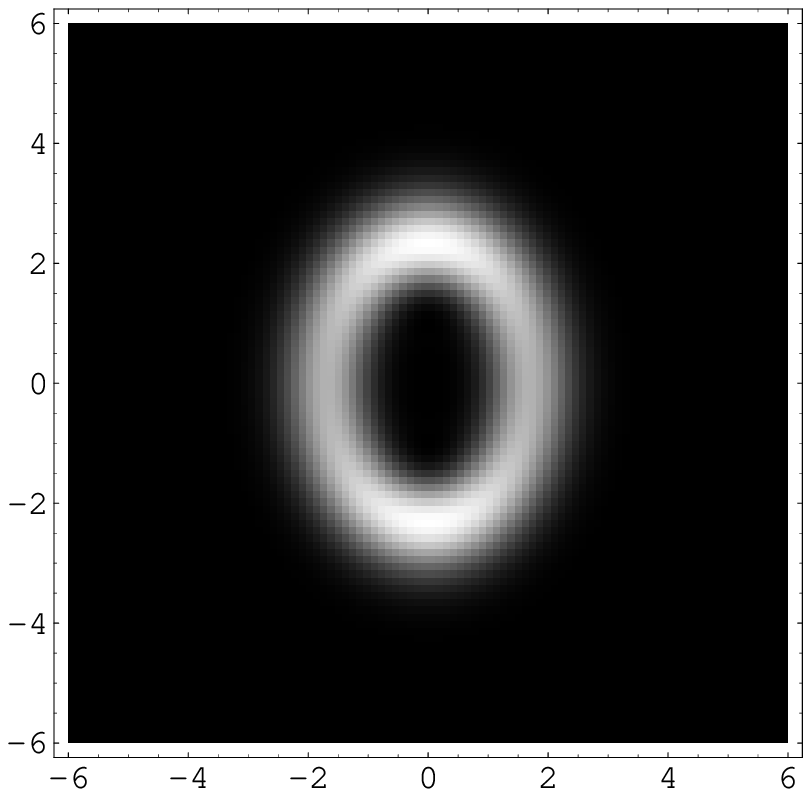,width=2in,height=2in}}
\vspace{0.25cm}
\psfig{file=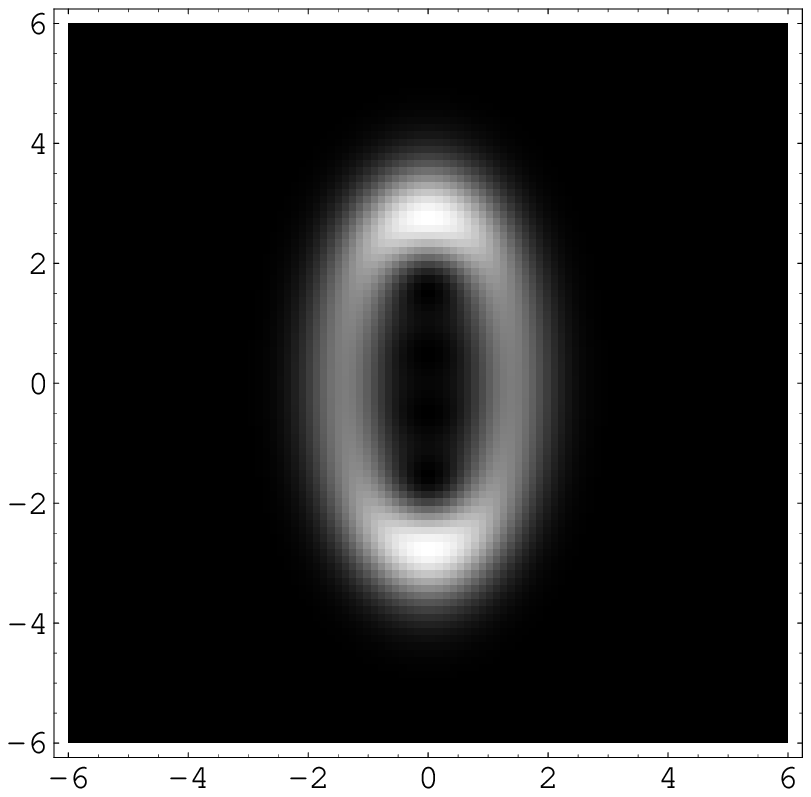,width=2in,height=2in}
\hspace{.25cm}
\psfig{file=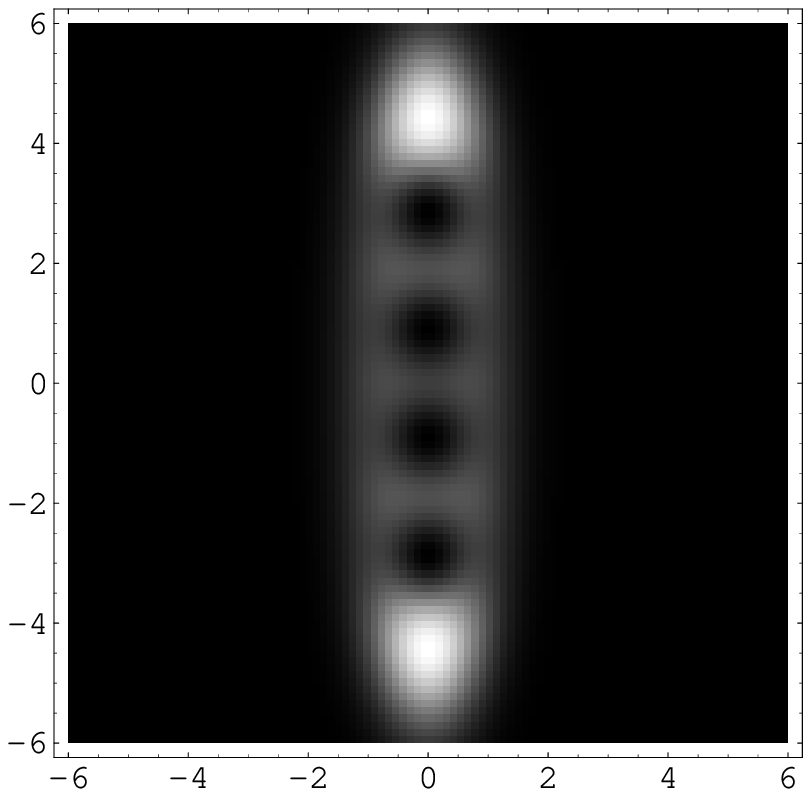,width=2in,height=2in}

\caption[]{Density of the 5$^{th}$ eigenstate at various anisotropies. The parameter $\lambda$ was
chosen as $0.05,0.35,.65,.95$ starting from top left figure. The zeros of the density are placed on the weak trapping axis,
at the roots of the 5$^{th}$ Hermite polynomial, as can be seen from the form of the wavefunction
Eq.[\ref{anisotropicwavefunctions}]
. Length is measured in units of oscillator length.}
\end{figure}

The expression for the single particle wavefunctions in the small anisotropy limit Eq.(\ref{anisotropicwavefunctions}),
is one of the central results of this paper. In the following sections, we will use these wavefunctions to discuss the
properties of a vortex lattice in an anisotropic rotating trap.

\section{Variational Wavefunction for the vortex lattice}

After solving the one particle problem exactly, we now consider a BEC containing many particles in a rotating anisotropic
trap. Here the system  is very well described by the Gross-Pitaevskii energy functional
\be
\label{gpenergy}
E =  \int dx \; dy \Psi^* \left[ \frac{1}{2} \left[(1+\eta) x^2+ (1-\eta) y^2 +p_x^2+p_y^2 \right] - \Omega L_z \right] \Psi
 +\frac{g}{2} \int dx \; dy |\Psi|^4.
\ee
In the fast rotation limit considered in this paper, the behavior of the cloud in the z direction can be easily described
by Thomas-Fermi approximation. The validity of this approximation has been experimentally checked by viewing the vortex
lattice from the side \cite{ho01,kav02}. Experimental results in an axisymmetric trap confirm that the vortices are only marginally bent along the z direction, as
predicted by the Thomas-Fermi approximation \cite{ech02}. We also expect the vortex lattice in an anisotropic trap to show the
same property and have straight vortices through the cloud. Therefore in the following discussion we will only consider
the Gross-Pitaevskii energy functional in two dimensions. This will allow us to use the results of the previous section
more directly.

In the small anisotropy limit introduced in the previous section, we have found that the LLL of the anisotropic trap only
has an overlap with the LLL of the axisymmetric trap. Thus any wavefunction in the lowest Landau level of the anisotropic
problem can be written as
\be
\Psi(x,y) = f(z) e^{\theta_y z^2} e^{-z \bar{z}} = g(z) e^{-z \bar{z}},
\ee
here both $f(z)$ and $g(z)$ are analytic functions of $z$. The zeros of the function $f(z)$ (or $g(z)$) correspond to the
positions of the  vortices in the BEC. Thus, a BEC containing a vortex lattice will correspond to an analytic function $f(z)$
that has a regular array of zeros. An analytic function that both has a regular array of zeros and that gives a normalizable
wavefunction  can be introduced  using the Jackobi theta function \cite{abr57,tka66,mho02}. We write down the following variational wavefunction,
which we will refer to as the variational vortex lattice wavefunction
\bea
\label{fullthetafunction}
\Psi(x,y) = C \Theta(\zeta e^{i \varphi},\tau) e^{-\frac{\pi}{2 \nu_0} (x \sin \varphi + y \cos \varphi)} \\ \nonumber
\times e^{\frac{\pi}{2 \nu_0} (x \sin \varphi + y \cos \varphi)} e^{\frac{\gamma+\theta_y}{2} (x^2-y^2+2ixy)}
e^{-\frac{x^2+y^2}{2}}.
\eea
Here $C$ is a normalization constant, $\varphi$ determines the angle of the lattice basis vectors with the $x$ axis.
We define the lattice basis vectors $a_1$ and $a_2$ as complex numbers in the $z=x+iy$ plane and $\nu_0$ is the volume of
the unit cell
\be
\nu_0= Im\{a_1^* a_2\}.
\ee
Then $\zeta$ and $\tau$ are defined as
\be
\zeta= z/|a_1|,
\ee
and
\be
\label{udefinition}
\tau = u + i v = a_2/a_1.
\ee
Finally $\gamma$ is a variational parameter that is in general complex, and is used to ensure that the wavefunction is
normalizable.

The Jackobi Theta function
\be
\Theta(\zeta,\tau) = -i \sum_{n=-\infty}^{\infty} (-1)^n e^{i \pi \tau (n+1/2)^2} e^{2 \pi i \zeta (n+1/2)}
\ee
is  a quasi-periodic function. The first line of the variational wavefunction Eq.[\ref{fullthetafunction}], the Theta
function with the exponential factor, is a periodic function in the complex plane \cite{cha85}.
Similar wavefunctions  have been used to
discuss vortex lattices in superconductors \cite{abr57}, superfluid He$^4$ \cite{tka66} and BEC in axisymmetric traps
\cite{ho01,mho02,mho03}.

The energy functional takes a simple form in the LLL as the projection of the angular momentum operator $L$ on to the LLL
is
\be
\int dx \; dy \Psi^*(z) L \Psi(z) = \int dx \; dy (r^2 - 1) |\Psi(z)|^2.
\ee
Thus, the Gross-Pitaevskii energy functional takes the form
\be\
\label{variationalenergy}
E = \delta \int dx \; dy |\Psi|^2 (x^2+y^2) + \frac{\eta}{2} \int dx \;dy |\Psi|^2 (x^2-y^2)
+\frac{g}{2} \int dx \; dy |\Psi|^4.
\ee

We minimize this energy functional with respect to the variational parameters in the wavefunction to understand the behavior
of a BEC containing a vortex lattice. The first property we are interested in is the  overall density profile of the
cloud.

Unlike a BEC in an axisymmetric trap which preserves its rotational symmetry and spreads out evenly in all directions, a
BEC in an  anisotropic trap will prefer to stretch along the weak axis of the trap. Even if the anisotropy of the potential
is small, if the gas is rotating close to the trapping frequencies the cloud will be stretched along the weak axis of the
potential.

To find the effect of the  anisotropy on the cloud profile, we can average out the vortices to write an 'averaged vortex'
wavefunction. This will correspond to replacing the periodic part of our variational wavefunction with a constant
\be
\Psi = C' e^{2\left[\frac{\theta_y+\gamma}{2} -\frac{1}{2}\right] x^2 + 2 \left[ \frac{\pi}{2 \nu_0}
- \frac{\theta_y+\gamma}{2} -\frac{1}{2} \right] y^2}.
\ee
In this approximation, the orientation of the vortex lattice with respect to the potential is not important, as one
can always choose the variational parameter $\gamma$ to compensate for the effect of the orientation of the lattice
on the averaged density profile. Furthermore, it is evident from the energy functional that cloud will be stretched
along the weak axis of the potential. We see that the averaged cloud density will be a
gaussian of the form.
\be
\Psi = C e^{-\frac{1}{2} \alpha_x^2 x^2- \frac{1}{2} \alpha_y^2 y^2}.
\ee

Doing the variational calculation, we find the half width of the cloud in both directions as
\bea
\alpha_x^{-1}&=& \frac{1}{2} \sqrt{\frac{(1-\lambda^2)^{1/4}}{(1+\lambda)} \sqrt{\frac{g N}{2 \pi \delta}}} \\
\alpha_y^{-1}&=& \frac{1}{2} \sqrt{\frac{(1-\lambda^2)^{1/4}}{(1-\lambda)} \sqrt{\frac{g N}{2 \pi \delta}}}
\eea
and the vortex density $\nu_0$ as
\be
\frac{\pi}{2 \nu_0} = 1- \frac{1}{(1-\lambda^2)^{1/4}} \sqrt{\frac{2 \pi \delta}{g N}}.
\ee
Here, $N$ is the total number of particles in the cloud and is related to the normalization constant $C$ as
\be
|C|^2 = \frac{\sqrt{1-(\theta_y+\gamma)} \sqrt{1+(\theta_y+\gamma)-\frac{\pi}{\nu_0}}}{\pi} N.
\ee
We see that as rotation frequency approaches critical rotation $\delta=\eta/2$, the cloud is stretched along the y
direction and becomes more one dimensional. The averaged vortex approximation breaks down  when the minor axis of the
cloud becomes as small as the distance between the vortices. The aspect ratio of the cloud $\frac{\alpha_y}{\alpha_x}$
is plotted in Fig.[2] as a function of rotation frequency. This aspect ratio can be experimentally measured in a set up
such as Ref.\cite{rps02}, without the experimental sophistication needed to image individual vortices.
It is worth noting that this aspect ratio is smaller than the
aspect ratio of the one particle ground state in the same trap.

\begin{figure}
\centerline{\psfig{file=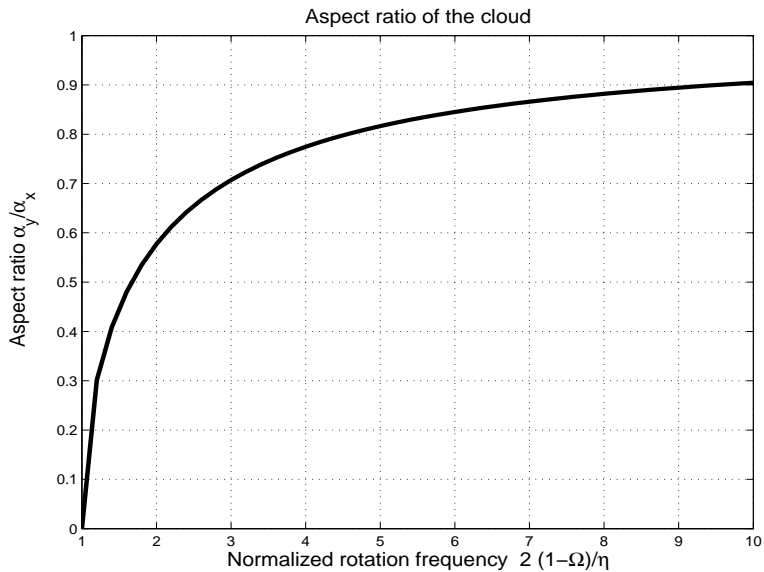,width=4in,height=3in}}
\caption[]{
The aspect ratio of the cloud is defined as the ratio of the  radii of the cloud in x direction and y direction.
Aspect ratio is plotted against $\lambda^{-1}$, which is the scaled rotation frequency. Here $\lambda^{-1}$ starts
from the critical value 1, and increases. As the rotation frequency decreases the aspect ratio becomes closer to 1
which is the value for an axisymmetric trap. }
\end{figure}

We note here that the result obtained in the averaged vortex approximation can  also be obtained by an appropriate
generalization of the formulation in Ref.\cite{cst03}, where an averaged velocity field is used as a hydrodynamic variable.
After calculating the vortex density and the density profile of the cloud, we consider the structure of the vortex lattice
in the next two sections.

\section{structure of the vortex lattice}

To be able to discuss the structure of the vortex lattice, we need to go beyond the  averaged vortex approximation
of the previous section. Thus, instead of  replacing the periodic part of the wavefunction by a constant, we need to
consider all the Fourier components. As the energy functional $\epsilon$ depends  only on the absolute value of the
wavefunction but not on the phase, we find it useful to write
\be
|\Psi(\vec{r})|^2 = C g(\vec{r}) e^{-\alpha_x^2 x^2 - \alpha_y^2 y^2}.
\ee
Here $g(\vec{r})$ is periodic with
\be
g(\vec{r}+\vec{R}) = g(\vec{r}),
\ee
where
\be
\vec{R}= n \vec{a}_1 + m \vec{a}_2
\ee
are the lattice vectors of the vortex lattice. Also, $\vec{a}_1$ and $\vec{a}_2$ are the basis vectors of the unit cell and $n$,$m$
are integers. This periodicity allows us to introduce a Fourier series representation for $g(\vec{r})$
\be
g(\vec{r})= g_0 \left[ 1 + \sum_{\vec{K}} \frac{g_{\vec{K}}}{g_0} e^{- i \vec{K} \cdot \vec{r}} \right].
\ee
Where $\vec{K}$ are the reciprocal lattice vectors
\be
\vec{K} =  n \vec{b}_1 + m \vec{b}_2,
\ee
with $n$,$m$ integers and the reciprocal lattice basis vectors given by
\bea
\vec{b}_1 &=& \hat{z} \times \vec{a}_1 \\ \nonumber
\vec{b}_2 &=& \vec{a}_2 \times \hat{z}.
\eea
To find the equilibrium vortex lattice configuration we need to minimize the energy functional with respect to the lattice
basis vectors. In doing this minimization, it is important to notice that if there are many  vortices in the cloud, there
will always be  a large number of vortices along both the minor and the major axes of the cloud
\bea
\label{manyvortexcondition}
\alpha_x \nu_0 &\ll& 1 \\ \nonumber
\alpha_y \nu_0 &\ll& 1.
\eea
Thus, for all reciprocal lattice vectors $\vec{K}$ except $\vec{K}=0$, we have
\bea
\alpha_x &\ll& |\vec{K}| \\ \nonumber
\alpha_y &\ll& |\vec{K}|
\eea

This gives us the important result that the contribution of the higher fourier components of the vortex lattice to the
first two terms in the energy functional are suppressed by the exponential factors $e^{K_x^2/\alpha_x^2}$ or
$e^{K_y^2/\alpha_y^2}$.  So, if the conditions Eq.(\ref{manyvortexcondition}) are satisfied, {\it i.e.} the cloud is
not extremely stretched, it is sufficient to consider the minimization of the interaction term in the energy functional
\be
\frac{g}{2} \int dx \; dy |\Psi|^4 = \frac{g}{2} C^2 \int dx \; dy e^{-2 \alpha_x^2 x^2 - 2 \alpha_y^2 y^2}
\left| 1 + \sum_{\vec{K}} \frac{g_{\vec{K}}}{g_0} e^{- i \vec{K} \cdot \vec{r}} \right|^2.
\ee
Once again in the multiplication of the fourier components any terms  that depend on $\vec{r}$ will be exponentially
small. Thus, the optimum lattice is found by the minimization of
\be
\sum_{\vec{K}} \frac{|g_{\vec{K}}|^2}{|g_0|^2}.
\ee
This, however, is exactly the minimization done for He$^4$ and BEC in axisymmetric traps \cite{mho02,tka66}. Resulting vortex lattice
is, as in the other cases, triangular. We then find that the vortex lattice of a BEC containing a large number of
vortices stays triangular even if the confining potential is anisotropic.

Physically, the robustness of the vortex lattice can be explained as follows. If there are enough
vortices in the cloud, the stress created by anisotropic confinement is completely screened and the structure
of the vortex lattice is determined locally by the interaction between the vortices. Unless the trap is not quadratic,
or the cloud is extremely stretched, the minimum energy configuration of the vortices will correspond to a triangular
lattice.

It is, however, instructive to calculate the corrections to the triangular lattice. The first correction to the
lattice will come from the change in the cloud's shape, not directly from the anisotropy of the trap. Thus, the deviation
from the triangular lattice will increase closer the rotation frequency gets to the weak trapping potential. To calculate
this deviation, we once again turn our attention to the interaction term in the energy functional.
\be
\epsilon_I = \frac{g}{2} C^2 \int dx \; dy e^{-2 \alpha_x^2 x^2 - 2 \alpha_y^2 y^2}
\left| 1 + \sum_{\vec{K}} \frac{g_{\vec{K}}}{g_0} e^{- i \vec{K} \cdot \vec{r}} \right|^2.
\ee
Now instead of considering terms that have no position dependence, we need to consider all the terms that have
position dependence only in the strong confining direction, {\it i.e.} the minor axis of the cloud. This is due to the
fact that as the cloud gets more and more stretched we get
\be
e^{-\alpha_x^2 x^2} \gg e^{-\alpha_y^2 y^2}.
\ee
Thus, to study the structure of the lattice we need to minimize
\be
I = \sum_{\vec{K},\vec{K'},K_y=K'_y} \frac{g_{\vec{K}} g_{\vec{K'}}}{g_0^2} e^{-\frac{(K_x-K'_x)^2}{8 \alpha_x}}.
\ee

Now, it is important to consider the orientation of the vortex lattice. In the next section we show that the lattice planes
prefer to orient themselves along the weak trapping direction, thus we choose the reciprocal lattice  basis vectors as
\bea
\vec{b}_1 &=& b_1 \hat{x} \\ \nonumber
\vec{b}_2 &=& b_1 (u \hat{x} + v \hat{y}).
\eea
The fourier components $g_{\vec{K}}$ can be calculated by expanding the $\theta$ function \cite{mho02,cha85}
 as a sum of Gaussians
\be
g_{\vec{K}}=(-1)^{m_1+m_2+m_1 m_2} e^{-\nu_0 \frac{|\vec{K}|^2}{8 \pi}} \sqrt{\frac{\nu_0}{2}},
\ee
for
\be
\vec{K} = m_1 \vec{b}_1 + m_2 \vec{b}_2.
\ee
Thus, we minimize
\be
I = \sum_{m_1,m_2,n_1} (-1)^{(m_1+1)(n_1-m_1)/2}  e^{-\frac{\pi}{2 v} \left[ (v m_1)^2 + (v n_1)^2
+ 2 (m_2 - u m_1)^2 - \frac{1}{\alpha_x \nu_0}(v (m_2 - n_2))^2 \right] }.
\ee
This minimization can be done numerically, as the exponential dependence causes a very rapid convergence of the sum.
We find that the triangular lattice becomes compressed in the weak  confinement direction as the rotation frequency
approaches confining frequency. The lattice planes, all of which are aligned in the weak confining direction, move apart
while keeping vortex density constant. This increases the density of vortices inside each plane, as can be expected.
We must remark here that this compression is a very small effect, even at $\alpha_x \nu_0 \sim 1$ where the theory
is expected to break down, the fractional change in the lattice constant is less then 0.5\%. It seems very unlikely that any
deviation from triangular lattice to be seen in an experiment.
As a function of $\alpha_x \nu_0$ this dependence is plotted in Figure 3.

We see that unless the cloud is extremely stretched so that there are very few vortices along the minor axis, the
deviation from the triangular lattice is small. Thus a BEC in equilibrium with a quadratic rotating potential
will always have triangular vortex lattice and would not show any structural phase transitions.


\begin{figure}
\centerline{\psfig{file=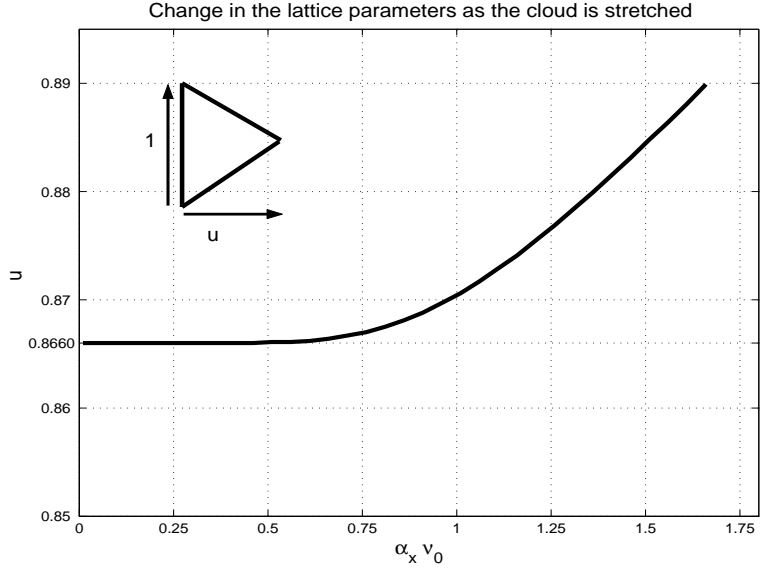,width=4in,height=3in}}
\caption[]{ The separation between the lattice planes $u$ (see Eq.(\ref{udefinition}))is plotted as a function of  $\alpha_x \nu_0$. This parameter
$\alpha_x \nu_0$ can be regarded as one over the number of vortices along the minor axis of the cloud. $u=\sqrt(3)/2$
corresponds to a triangular lattice, and we see that the deviation from the triangular lattice remains less than 0.5 \%
even around $\alpha_x \nu_0 \sim 1$, where the theory presented here should break down.
 }
\end{figure}

\section{ Orientation of the vortex lattice}
Upon the formation of a vortex lattice in an axisymmetric trap, the full rotational symmetry of the system is
reduced to the point symmetry group of the vortex lattice. For a triangular lattice this group is $C_6$. The angle
of the vortex lattice with respect to a fixed direction in the rotating frame is, however, arbitrary. This fact is
experimentally confirmed by viewing the vortex lattice from the side in JILA experiments \cite{ech02}.

For an anisotropic trap, there is already a fixed direction in the rotating frame that breaks the full rotational
symmetry. Energy of the vortex lattice depends on the orientation of the lattice planes with respect to the  anisotropy
of the cloud. Using our trial wavefunction we calculate this energy for an arbitrarily oriented triangular vortex lattice.

The dependence of the energy on the orientation of the vortex lattice will be due to the only non-rotationally invariant
term in the energy functional
\be
\epsilon_2 = \frac{\eta}{2} \int dx \; dy \; |\Psi|^2 (x^2 - y^2).
\ee
Again we write the cloud density from our variational wavefunction as
\be
|\Psi|^2 = C g(\vec{r}) e^{-\alpha_x^2 x^2-\alpha_y^2 y^2},
\ee
with $g(\vec{r})$ a periodic function
\be
g(\vec{r}) = g(\vec{R}+\vec{r}),
\ee
and $\vec{R}$ the lattice vectors of the vortex lattice
\be
\vec{R} = n a_1 + m a_2.
\ee
For a triangular lattice that makes an angle $\varphi$ with the minor axis of the cloud
\bea
\vec{a}_1 &=&  a ( \cos(\varphi) \hat{x} + \sin(\varphi) \hat{y} ) \\ \nonumber
\vec{a}_2 &=&  a \left( \frac{1}{2} (\cos(\varphi) \hat{x} + \sin(\varphi) \hat{y} ) + \frac{\sqrt{3}}{2} ( - \sin(\varphi) \hat{x}
+ \cos(\varphi) \hat{y}) \right),
\eea
and the vortex density is
\be
\nu_0 = \frac{ \sqrt{3} a^2}{4}.
\ee

By Fourier transforming the  density and evaluating the resulting gaussian integrals we calculate the anisotropy term
in the energy functional as
\bea
\epsilon_2 &=& \frac{\eta}{4} C^2 g_0 \frac{\pi}{\sqrt{\alpha_x \alpha_y}} \sum_{n,m} \left[ \left( \frac{1}{\alpha_x}
- \frac{1}{\alpha_y} \right) - \left( \frac{(\vec{K}_{nm} \cdot \hat{x})^2}{2 \alpha_x^2} - \frac{(\vec{K}_{nm} \cdot \hat{y})^2}{2 \alpha_y^2}
\right) \right] \\ \nonumber
&\times& \frac{g_{\vec{K}_{nm}}}{g_0} e^{-\frac{(\vec{K}_{nm} \cdot \hat{x})^2}{4 \alpha_x}} e^{-\frac{(\vec{K}_{nm} \cdot \hat{y})^2}{2 \alpha_y}}.
\eea
Here the sum $n$,$m$ is over all positive integers and
\be
\vec{K}_{nm} =  n \vec{b}_1 + m \vec{b}_2,
\ee
with $\vec{b}_1$,$\vec{b}_2$ are the reciprocal lattice vectors
\bea
\vec{b}_1 &=& \frac{ 4 \pi}{\sqrt{3} a} \left( \vec{a}_2 \times \hat{z} \right) \\ \nonumber
\vec{b}_2 &=& \frac{ 4 \pi}{\sqrt{3} a} \left( \hat{z} \times \vec{a}_1 \right).
\eea

Now the exponential factors get very small as $n$,$m$ increase, thus by summing over the closest six $\vec{K}$ points to
the origin we obtain the first contribution to  the orientational energy of the vortex lattice as
\bea
\epsilon_{\varphi} = &-& \frac{\eta}{\alpha_y^2} N \frac{\pi}{\sqrt{3} \nu_0} e^{-\frac{\sqrt{3}}{2} \pi} \Big[
\sin^2(\varphi-\frac{\pi}{6}) e^{- \frac{\pi}{\sqrt{3} \alpha_y \nu_0} \sin^2(\varphi-\frac{\pi}{6})} \\ \nonumber
&+&
\cos^2(\varphi) e^{- \frac{\pi}{\sqrt{3} \alpha_y \nu_0} \cos^2(\varphi)}+
\sin^2(\varphi+\frac{\pi}{6}) e^{- \frac{\pi}{\sqrt{3} \alpha_y \nu_0} \sin^2(\varphi+\frac{\pi}{6})} \Big].
\eea

We see that the lattice planes prefer orienting themselves with the weak anisotropy direction. The physical reason for
this alignment is that the maximum amount of angular momentum can be gained when there are more vortices at the minimum
of the confining potential. The situation is then similar to having the first vortices at the center of a BEC when vortices
are initially formed.

This pinning of the orientation of the vortex lattice should be the first noticeable effect of the anisotropy of the
cloud on the vortex lattice. This effect was first pointed out for small number of vortices in Ref.\cite{kav02}.
 The orientation of the vortex lattice planes can easily be checked experimentally by viewing
the BEC from the side as done in the recent JILA experiments.

\section{ BEC with a small number of vortices}

After discussing a BEC containing a large number of vortices, we
turn our attention to the case where there are only a few vortices
in the condensate. Although, at first it may seem that the fast
rotation, {\it i.e.} LLL, assumption made through out the paper would
prevent the the discussion of this limit, this is not the case.
The number of vortices in the condensate is determined not only by
the closeness of the rotation frequency to the trapping frequency,
but also the  strength of the interactions between the particles
in the condensate \cite{kmp00,bro99}. A non-interacting gas, for example, would not
exhibit any vortex states, no matter how fast it is rotating.

The relevant energy scales for determining the number of vortices
are the rotation frequency $\delta$, the strength of the
interactions $g N$, where $N$ is the total number of particles in
the condensate, and for an anisotropic trap the anisotropy
parameter $\eta$. As the ratio $\frac{g N}{\delta}$ increases from
$0$, vortices will start appearing in the cloud. So if the
interactions are weak enough or the number of particles in the
condensate are small, a BEC which has a small number of vortices
but which is rotating fast enough to be entirely in the LLL is
possible. Also any vortex state wavefunctions obtained in the LLL
should be good variational wavefunctions for slower rotating
vortex states as LLL wavefunctions adequately represent the
analytic properties of all the vortex states.

In the very first experiments that demonstrated the presence of
vortices in a BEC, small numbers of vortices were observed \cite{mah99,mcw00,hhh02}. These
vortices were found to form very regular configurations. We
investigated whether the regularity of these arrays would be
affected if the confining potential were made anisotropic. By
numerical calculation, we found that the regular configurations of
the vortices are sensitive to the anisotropy of the confining
potential. Under anisotropy, the vortex configurations first
oriented themselves to accomodate the maximum number of vortices
on the weak trapping axis and then, as the anisotropy is
increased, became stretched along the same axis. Our approach is similiar to Ref.\cite{lnf01}, and
we obtain similar results with our approximate wavefunctions Eq.(\ref{anisotropicwavefunctions}). This can
be regarded as a success of the approximation scheme introduced in Section II.

To determine the vortex states, we numerically solved the
Gross-Pitaevskii equation using a truncated set of single particle
wavefunctions found in Section II. We expand the condensate
wavefunction $\psi(r)$ as
\be
\psi(r) = \sum_{n=0}^{n=n_{cut}} c_n \Psi_n(r).
\ee
Using this expansion in the expression for the Gross-Pitaevskii energy functional we obtain
\be
\epsilon = \sum_n \epsilon_n |c_n|^2 + \sum_{n,m,p,q} R_{nmpq} c^*_n c^*_m c_p c_q.
\ee
Here $\epsilon_n$ are the eigenvalues of the n$^{th}$ single particle state calculated in Section II, and
\be
R_{nmpq} = \int dx \; dy \Psi^*_n \Psi^*_m \Psi_p \Psi_q,
\ee
are calculated using the generating function for Hermite polynomials as
\bea
R_{nmpq} &=&  \frac{2^{-(n+m+p+q)/2} (\gamma_y+\gamma_y^{-1})^{-2}}{\pi \sqrt{n! m! p! q!}} |\theta_y|^{(n+m+p+q)/2} \\ \nonumber
&\times&(\partial_r)^n (\partial_s)^m (\partial_t)^p (\partial_u)^q e^{-\frac{1}{2} [(r-s)^2+(t-u)^2] }
e^{\frac{1}{|\theta_y|} (r+s)(t+u)}  \Big|_{r,s,t,u=0}.
\eea
We numerically minimized the energy functional by varying $c_n$ and plotted the density of some of the representitive
vortex configurations in Fig.[3]. In this minimization the parameters of the Hamiltonian were chosen so that the cutoff imposed
on the wavefunctions did not significantly alter the results of the minimization.

As a simple example to demonstrate the effect of anisotropy
consider a BEC with four vortices (See Fig.[4]). If this state was created in
an axisymmetric trap, the vortices form a square,
with the origin as the square's center. If an anisotropy is
introduced, we find that, first two of the vortices align themselves
with the weak trapping direction and then  move a little bit
further away from the origin. The other two approach the weak
trapping axis, thus, turning the square into a rhombus.
As anisotropy is increased, the rhombus becomes
more compressed, while conserving its form. After a critical value
of the anisotropy, the vortices, instead of forming a rhombus,
become aligned in the weak trapping direction. This critical
value, however depends on the ratio of interaction strength to the
rotation frequency $\frac{g N}{\delta}$.

\begin{figure}
\label{4vorfig}
\centerline{\psfig{file=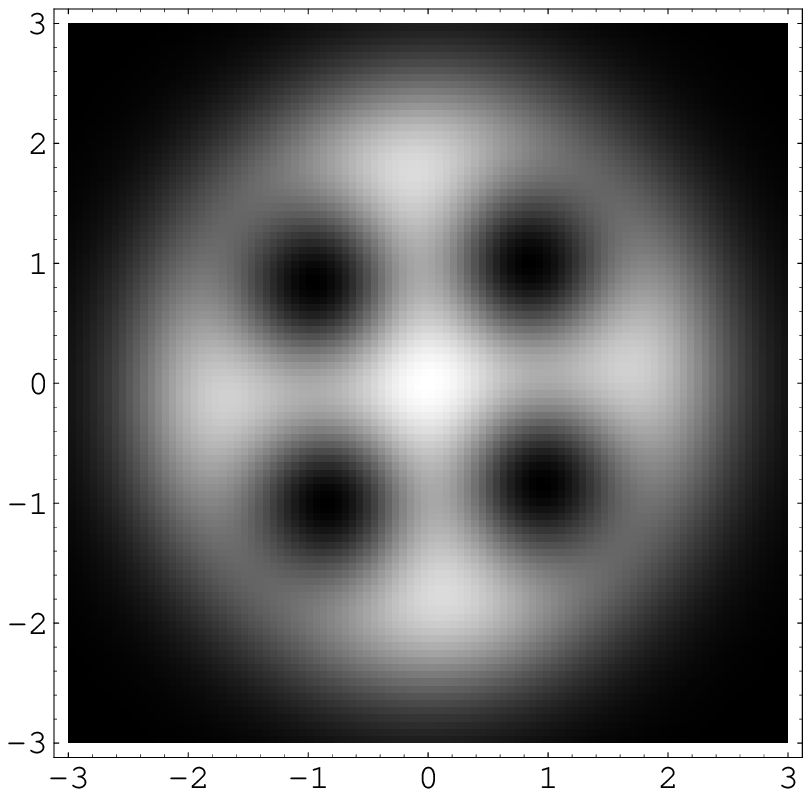,width=2in,height=2in}
\hspace{.25cm}
\psfig{file=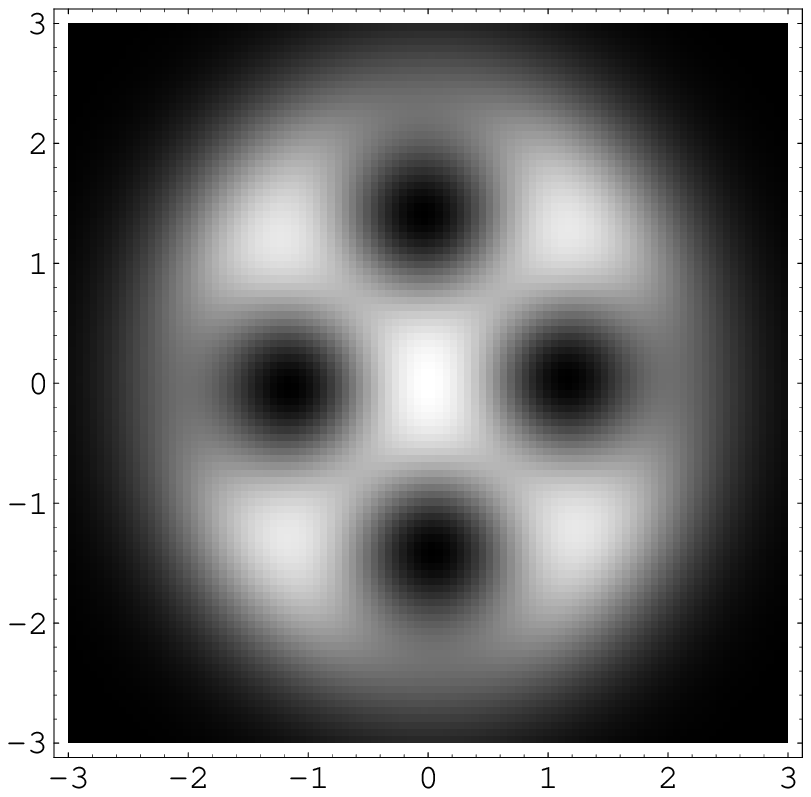,width=2in,height=2in}}
\vspace{0.25cm}
\psfig{file=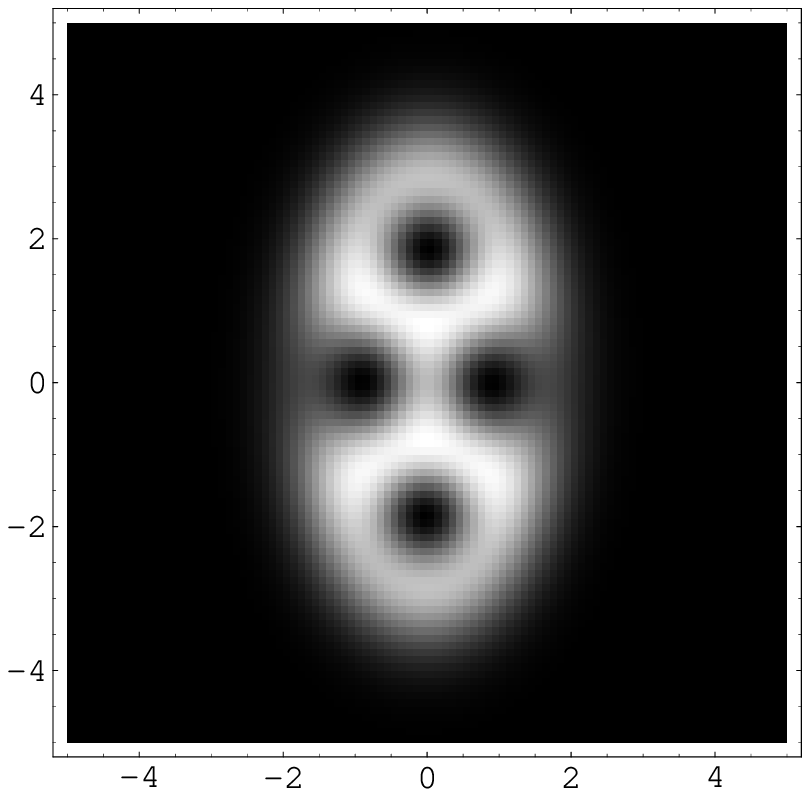,width=2in,height=2in}
\hspace{.25cm}
\psfig{file=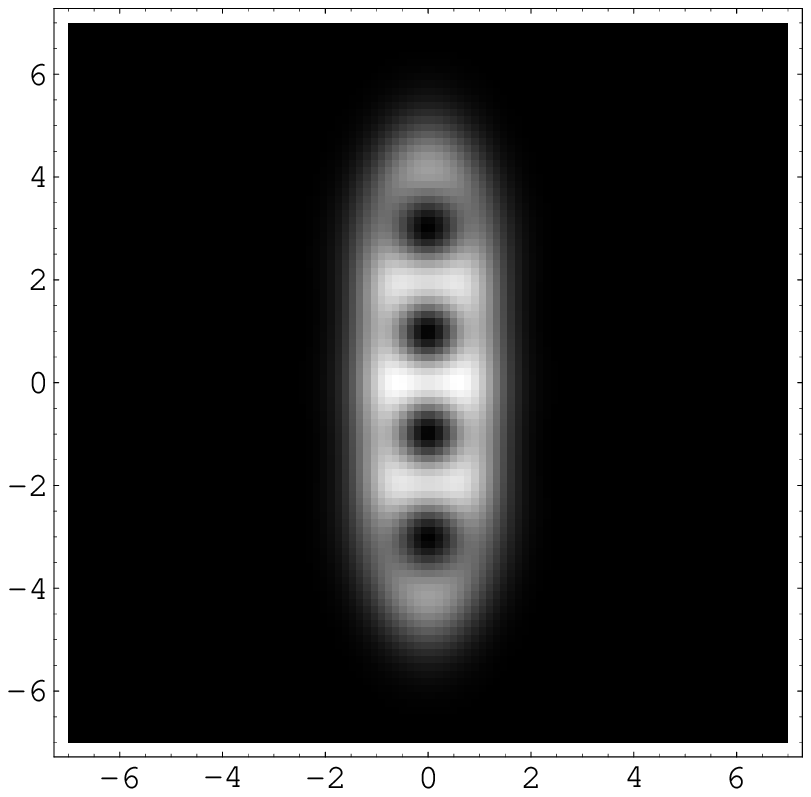,width=2in,height=2in}

\caption[]{Density of various 4 vortex states obtained by numerical calculation. The parameter $\lambda$ was
chosen as $0.05,0.25,.85,.90$ starting from top left figure. Interactions were chosen to result in 4 vortex states
at each different anisotropy. Length is measured in units of oscillator length.}
\end{figure}

A similar scenario plays out for vortex states with higher numbers
of vortices in them. The initial highly regular configuration
becomes  compressed in the strong confinement direction and at
some critical values of the anisotropy one or more vortices jump
to the weak trapping axis. In the situation of extreme anisotropy,
all the vortices become aligned on the weak trapping axis, forming
a one dimensional vortex lattice. It is important to notice that,
the behavior obtained in numerical calculations is in agreement
with the analytical calculation made in the previous sections.

In  Fig.[4], we display some of the obtained vortex states, by
plotting the two dimensional particle density. We have calculated
states with up to 8 vortices in them and believe that these
configurations can be experimentally observed by imposing a
rotating anisotropic potential on the few vortex states obtained
in experiments \cite{mcw00,hhh02}.

\section{Conclusion}

We studied the properties of vortex arrays in a BEC that is
confined by a rotating anisotropic potential. All our calculations
were made in the experimentally accessible limit of fast rotation,
however we argued that results obtained under this assumption
should also apply to slower rotating condensates.

To facilitate the discussion of a condensate in an anisotropic
rotating potential we first solved the one particle problem
exactly and obtained analytical expressions for the single
particle wavefunctions.

We obtained a variational wavefunction for a BEC containing a
vortex lattice using the results of the single particle problem.
This variational wavefunction enabled us to calculate the density
profile of the cloud, which could be easily measured
experimentally. Also, we found that even under anisotropic
confinement the vortex lattice stays triangular. However
orientation of the lattice planes is controlled by the external
potential.

For a condensate with a small number of vortices, we numerically
calculated the wavefunction using a truncated basis of single
particle eigenstates. We showed that the regularity of the vortex
configurations observed in an axisymmetric trap are disturbed by
the anisotropy of the confinement. We found that at high
anisotropy vortices form one dimensional arrays. We believe these
new vortex configurations can be experimentally obtained.

{\acknowledgements
We would like to thank Tin-Lun Ho (Ohio State) for his guidance and support,  and Erich Mueller (Cornell) for many
useful discussions.}

\end{document}